\documentclass[apjl]{emulateapj-rtx4}
\def\psr{{\rm PSR J0437$-$4715}}
\def\psrname{{\rm J0437$-$4715}}
\def\Msun{{\rm $ M _{\odot} $}}

\def\PSRFITS{{\tt PSRFITS}}
\def\PSRCHIVE{{\tt PSRCHIVE}}
\def\DSPSR{{\tt DSPSR}}
\def\PRESTO{{\tt PRESTO}}

\def\Curtin{$^{1}$}
\def\CAASTRO{$^{2}$}
\def\RRI{$^{3}$}
\def\Swin{$^{4}$}
\def\SKASA{$^{5}$}
\def\Rhodes{$^{6}$}
\def\ASU{$^{7}$}
\def\ANU{$^{8}$}
\def\Haystack{$^{9}$}

\def\MIT{$^{11}$}
\def\CfA{$^{12}$}
\def\UW{$^{13}$}
\def\Victoria{$^{14}$}
\def\UWisc{$^{15}$}
\def\UMichigan{$^{16}$}
\def\CASS{$^{17}$}
\def\Tata{$^{18}$}
\def\NRAO{$^{19}$}
\def\UMelbourne{$^{20}$}

\newcommand{\be}{\begin{eqnarray}}
\newcommand{\ee}{\end{eqnarray}}

\def\la{\mbox{\raisebox{-0.1ex}{$\scriptscriptstyle \stackrel{<}{\sim}$\,}}}
\def\ga{\mbox{\raisebox{-0.1ex}{$\scriptscriptstyle \stackrel{>}{\sim}$\,}}}
\newcommand{\nuobs}{\mbox{${\rm \nu _{obs} }$\,}}
\newcommand{\Viss}{\mbox{${V _{\rm iss} }$\,}}
\newcommand{\Aiss}{\mbox{${A _{\rm ISS} }$\,}}
\newcommand{\Vmu}{\mbox{${V _{\rm \mu} }$\,}}
\newcommand{\Vbinm}{\mbox{${v _{\rm bin} }$\,}}
\newcommand{\Vbin}{\mbox{${v _{\rm bin,\perp} }$\,}}
\newcommand{\Vearth}{\mbox{${v _{\rm earth,\perp} }$\,}}
\newcommand{\Vscreen}{\mbox{${v _{\rm screen} }$\,}}
\newcommand{\nud}{\mbox{${\nu _{\rm d} }$\,}}
\newcommand{\nuc}{\mbox{${\nu _{\rm c} }$\,}}
\newcommand{\nudc}{\mbox{${\nu _{\rm d _{c} } }$\,}}
\newcommand{\dtnu}{\mbox{${d t / d \nu }$\,}}
\newcommand{\tauiss}{\mbox{${\tau _{\rm iss} }$\,}}
\newcommand{\nudmwa}{\mbox{${\nu _{\rm d,mwa} }$\,}}
\newcommand{\tauissmwa}{\mbox{${\tau _{\rm iss,mwa} }$\,}}
\newcommand{\driftunit}{${\rm s \ MHz ^ {-1}}$\,}
\newcommand{\cnsqunits}{${\rm m ^{-20/3}}$\,}
\newcommand{\smunits}{${\rm kpc\,m ^{-20/3}}$\,}
\newcommand{\dmu}{${\rm pc \ cm ^ {-3}}$\,}
\newcommand{\velu}{${\rm km \ s ^ {-1}}$\,}
\newcommand{\thetadiff}{\mbox{${{\rm \theta _{diff}}}$\,}}
\newcommand{\thetaref}{\mbox{${{\rm \theta _{ref}}}$\,}}
\newcommand{\Tobs}{\mbox{${T _{\rm obs}}$\,}}
\newcommand{\Dos}{\mbox{${D _{\rm os}}$\,}}
\newcommand{\Dps}{\mbox{${D _{\rm ps}}$\,}}
\newcommand{\Dpsr}{\mbox{${D _{\rm psr}}$\,}}
\newcommand{\thetaHtau}{\mbox{${\theta_{\mathrm H}/\theta_{\tau}}$}}
\newcommand{\Cnsq}{\mbox{${ \overline { C _{\rm n} ^2 } }$\,}}
\newcommand{\Cnldl}{\mbox{${ { C _{\rm n} ^2 (l)\,dl} }$\,}}

\shortauthors{Bhat et al.}
\shorttitle{ \psr\ with the Murchison Widefield Array}
\begin{document}
\title{The low-frequency characteristics of  \psr\ observed with the Murchison Widefield Array}
\medskip
\author{
N. ~D.~R.~Bhat\Curtin$^,$\CAASTRO,
S.~M.~Ord\Curtin$^,$\CAASTRO,
S.~E.~Tremblay\Curtin$^,$\CAASTRO,
S.~J.~Tingay\Curtin$^,$\CAASTRO, 
A.~A.~Deshpande\RRI, 
W.~van~Straten\Swin$^,$\CAASTRO,
S.~Oronsaye\Curtin$^,$\CAASTRO,
G.~Bernardi\SKASA$^,$\Rhodes$^,$\CfA, 
J.~D.~Bowman\ASU, 
F.~Briggs\ANU$^,$\CAASTRO,
R.~J.~Cappallo\Haystack, 
B.~E.~Corey\Haystack, 
D.~Emrich\Curtin,
R.~Goeke\MIT,
L.~J.~Greenhill\CfA,
B.~J.~Hazelton\UW, 
J.~N.~Hewitt\MIT, 
M.~Johnston-Hollitt\Victoria,
D.~L.~Kaplan\UWisc, 
J.~C.~Kasper\UMichigan$^,$\CfA, 
E.~Kratzenberg\Haystack, 
C.~J.~Lonsdale\Haystack, 
M.~J.~Lynch\Curtin, 
S.~R.~McWhirter\Haystack,
D.~A.~Mitchell\CASS$^,$\CAASTRO, 
M.~F.~Morales\UW, 
E.~Morgan\MIT, 
D.~Oberoi\Tata, 
T.~Prabu\RRI, 
A.~E.~E.~Rogers\Haystack, 
D.~A.~Roshi\NRAO, 
N.~Udaya~Shankar\RRI, 
K.~S.~Srivani\RRI, 
R.~Subrahmanyan\RRI$^,$\CAASTRO, 
M.~Waterson\Curtin$^,$\ANU,
R.~B.~Wayth\Curtin$^,$\CAASTRO, 
R.~L.~Webster\UMelbourne$^,$\CAASTRO, 
A.~R.~Whitney\Haystack, 
A.~Williams\Curtin, 
C.~L.~Williams\MIT}
\affil{
$^{1}$International Centre for Radio Astronomy Research, Curtin University, Bentley, WA 6102, Australia\\
$^{2}$ARC Centre of Excellence for All-sky Astrophysics (CAASTRO)\\
$^{3}$Raman Research Institute, Bangalore 560080, India\\
$^{4}$Centre for Astrophysics and Supercomputing, Swinburne University, Hawthorn, Victoria 3122, Australia\\
$^{5}$Square Kilometre Array South Africa, 3rd Floor, The Park, Park Road, Pinelands, 7405, South Africa\\
$^{6}$Department of Physics and Electronics, Rhodes University, PO Box 94, Grahamstown, 6140, South Africa\\
$^{7}$School of Earth and Space Exploration, Arizona State University, Tempe, AZ 85287, USA\\
$^{8}$Research School of Astronomy and Astrophysics, Australian National University, Canberra, ACT 2611, Australia\\
$^{9}$MIT Haystack Observatory, Westford, MA 01886, USA\\
$^{11}$Kavli Institute for Astrophysics and Space Research, Massachusetts Institute of Technology, Cambridge, MA 02139, USA\\
$^{12}$Harvard-Smithsonian Center for Astrophysics, Cambridge, MA 02138, USA\\
$^{13}$Department of Physics, University of Washington, Seattle, WA 98195, USA\\
$^{14}$School of Chemical \& Physical Sciences, Victoria University of Wellington, Wellington 6140, New Zealand\\
$^{15}$Department of Physics, University of Wisconsin--Milwaukee, Milwaukee, WI 53201, USA\\
$^{16}$Department of Atmospheric, Oceanic and Space Sciences, University of Michigan, Ann Arbor, MI 48109, USA\\
$^{17}$CSIRO Astronomy and Space Science, Marsfield, NSW 2122, Australia\\
$^{18}$National Centre for Radio Astrophysics, Tata Institute for Fundamental Research, Pune 411007, India\\
$^{19}$National Radio Astronomy Observatory, Charlottesville and Greenbank, USA\\
$^{20}$School of Physics, The University of Melbourne, Parkville, VIC 3010, Australia
}
\medskip
\begin{abstract}
We report on the detection of the millisecond pulsar \psr\ with the Murchison Widefield Array 
(MWA) at a frequency of 192 MHz. 
Our observations show rapid modulations of pulse intensity in time and frequency that arise from 
diffractive scintillation effects in the interstellar medium (ISM), as well as prominent drifts of intensity 
maxima in the time-frequency plane that arise from refractive effects.  
Our analysis suggests that the scattering screen is located at a distance of $\sim$80-120 pc from the Sun, 
in disagreement with a recent claim that the screen is closer ($\sim$10 pc).   
Comparisons with higher frequency data from Parkes reveals a dramatic evolution of the pulse 
profile with frequency, with the outer conal emission becoming comparable in strength to 
that from the core and inner conal regions. 
As well as demonstrating high time resolution science capabilities currently possible with the MWA, 
our observations underscore the potential to conduct low-frequency investigations of timing-array 
millisecond pulsars, which may lead to increased sensitivity for the detection of  nanoHertz 
gravitational waves via the accurate characterisation of ISM effects.  
\end{abstract}

\section{Introduction} \label{s:intro}

\psr, the closest and brightest millisecond pulsar (MSP), has been intensely studied ever since its discovery 
\citep{johnston+1993}. 
This 5.75-ms pulsar in a nearly-circular 5.74-day orbit with a low-mass ($\approx$0.24 \Msun) helium 
white-dwarf  companion has also been detected at optical, ultraviolet, X-ray and $\gamma$-ray wavelengths 
\citep{bell+1993,kargaltsev+2004,zavlin+2002,abdo+2013}.
It has the lowest dispersion measure (DM) of all known MSPs (2.65 \dmu) and is located at a distance 
of $157\pm3$ pc \citep{verbiest+2008,deller+2008}. 
\psr\ is among the most important objects for long-term high-precision timing applications and holds 
the record for timing precision over a decadal 
time span (rms timing residual=199 ns; \citet{verbiest+2008}). 
Consequently, it is amongst the most promising objects for pulsar timing-array experiments that aim to detect 
nanoHertz gravitational waves via high-precision timing.
It is a prime target for the ongoing Parkes pulsar timing array (PPTA) project \citep{manchester+2013}, and later this decade, 
the South African MeerKAT may also be used for its timing observations. 

Its proximity  makes \psr\ an important object for probing the local ISM, which is known to harbour large-scale 
features such as the Local Bubble \citep{snowden+1990,bhat+1998,ne2001,spangler2009}. 
Given its low DM and a location well below the Galactic plane ($b\approx-42^{\circ}$), it is likely to 
be relatively weakly scattered, and therefore observations at the MWA frequencies (\la 300 MHz) are 
most ideal to characterise the ISM along its line-of-sight. 
However, the reported scintillation work in the published literature is limited, and was undertaken at 
higher frequencies \citep{johnston+1998,gg2000,gwinn2006}. 
Even though scattering and refraction are generally expected to be negligible at frequencies 
\ga1 GHz, where high-precision timing is currently achievable, the PPTA data show significant  DM 
variations, $\sim10^{-3}$ \dmu over 
time spans  $\sim$5-10 yr \citep{keith+2013}, corresponding to timing perturbations $\sim$1-2 $\mu$s.
The dispersive delays within the MWA band will be 1-2 orders of magnitude larger than 
those at $\sim$700 MHz, the lowest frequency of PPTA observations.
Low-frequency investigations are therefore particularly important in order to characterise the 
ISM effects toward this pulsar. 

\psr\ also has a remarkable pulse profile, comprising multiple overlapping 
components, with emission extending out to nearly 85\% of the rotation period 
\citep{johnston+1993,mj1995,yan+2011}. 
It is also very well studied in polarisation \citep[e.g.][]{mj1995,navarro+1997}, and is an important 
polarimetric calibrator for timing-array experiments and high-precision pulsar polarimetry \citep[e.g.][]{vanstraten2013}.

While the PPTA observations are made at frequencies $\ga$700 MHz \citep{manchester+2013}, 
the pulsar has previously been detected at lower frequencies, down to 76 MHz \citep{mcconnell+1996}, 
and also at 151 and 330 MHz \citep{issur2002,vivek+1998}, though the latter two used instruments of 
single polarisation. 
Despite their limited time resolution and signal-to-noise, a strong frequency evolution of the pulse profile 
was hinted at by these early observations. 

In this Letter, we report on the detection of pulsar \psrname\ made with the MWA.
Our observations enable the lowest-frequency scintillation studies ever made of this pulsar, 
and reveal the signatures of both diffractive and refractive scintillation effects. 
Details of our observations and data processing are summarised in \S~\ref{s:obs}, discussion of results 
in \S~\ref{s:res}, and future prospects are outlined in \S~\ref{s:conc}. 
 
\begin{figure}[t]
\epsscale{1.0}
\includegraphics[width=3.5cm,angle=270]{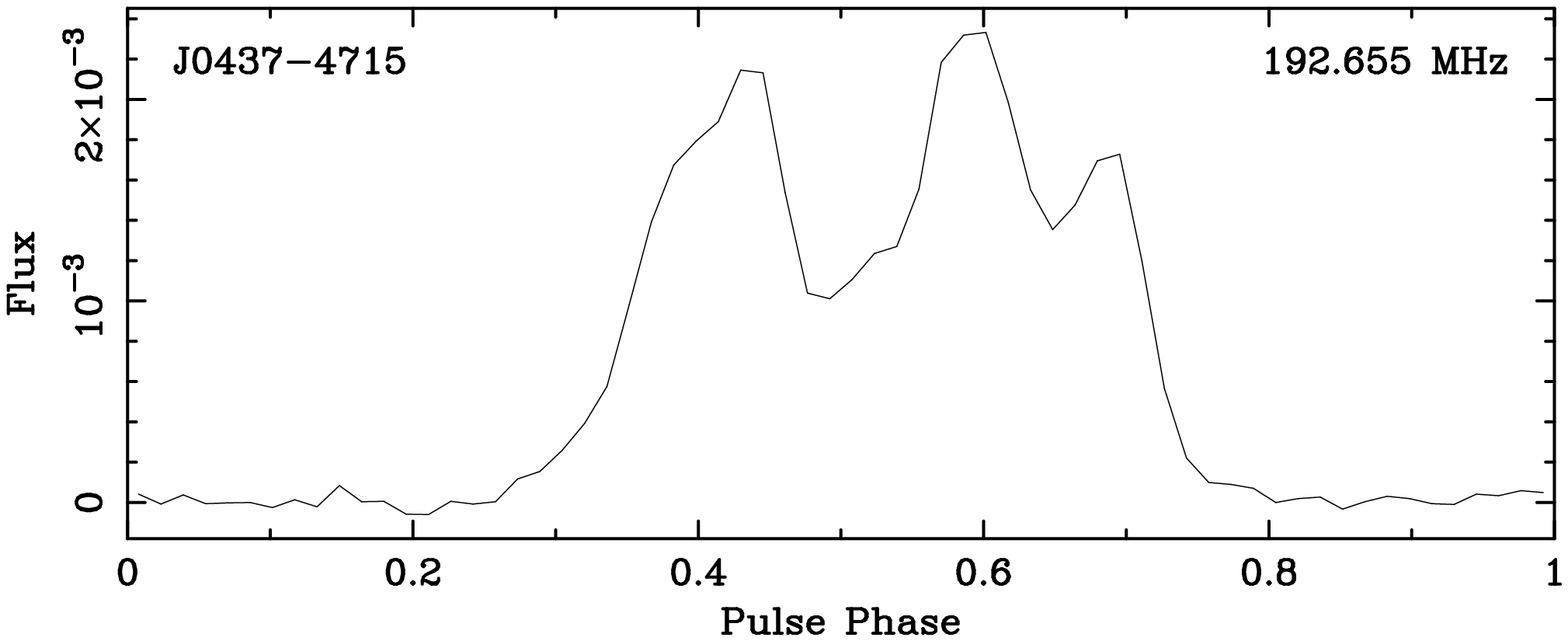}
\includegraphics[width=7cm,angle=270]{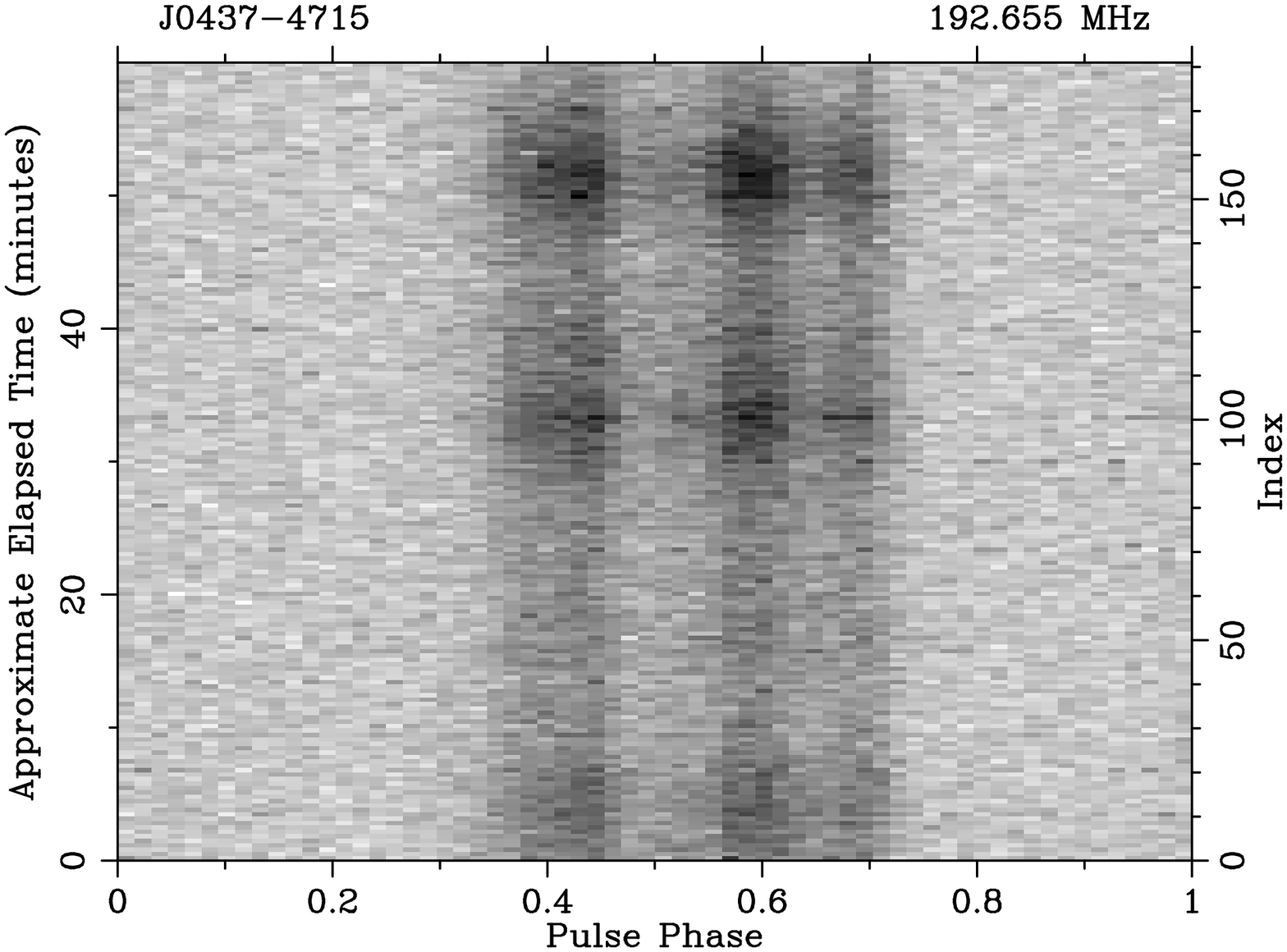}
\caption{Detection of \psr\ with the MWA. Integrated pulse profile (top), and 
the pulse strength versus time and pulse phase (bottom), for the 1 hr of observation 
centred at 192.6 MHz over a bandwidth of 15.36 MHz (``index" on right hand side 
of bottom panel refers to 20-second sub-integration). Data are de-dispersed, 
and the residual dispersive smearing within 10-kHz channels is $\approx$35 $\mu$s, smaller 
than the 100 $\mu$s time resolution.}
\label{fig:presto}
\end{figure}

\section{Observations and Data Processing}\label{s:obs}

The MWA is a newly operational low-frequency array that comprises 128 tiles  operating from 80 to 
300 MHz \citep{tingay+2013}. While originally 
conceived primarily as an imaging instrument
 \citep{lonsdale+2009}, 
 the array is equipped with a high time resolution data 
recorder -- the voltage capture system (VCS), to enable time-domain science applications 
such as observations of pulsars and fast radio bursts. Implementation details of the VCS will be 
described elsewhere (Tremblay et al. in prep.). This functionality allows recording up to $24\times1.28\,{\rm MHz}$ from 
all 128 tiles 
(both polarizations). These 24 coarse channels are further sub-divided into 10-kHz fine channels, 
resulting in a native time resolution of 100 $\mu$s. 
Recording of $12\times1.28\,{\rm MHz}$ is currently possible to provide a net bandwidth of 15.36 MHz, with 
the full-bandwidth capability anticipated later in 2014. 
Commissioning observations to date have resulted in the detection of eight pulsars including the Crab 
and \psr. 

Observations of \psr\ were made on two separate occasions, on 25 September 
2013 and 13 December 2013, and the data were recorded over a $12\times1.28\,{\rm MHz}$ 
bandwidth centred at 192.6 MHz. 
For each observation, the raw voltage data from VCS, at a rate of $328~{\rm MB\,s^{-1}}$ 
per 1.28 MHz coarse channel (i.e. 4-bit sampling), were recorded onto disks for a duration of 1 hr, 
resulting in an aggregate 13.5 TB of data per observation. 
Owing to a temporary limitation with our data recording,
only 88 of the $128\times$10-kHz  channels (in each coarse channel) were written to the disks. 
These data were then processed to form an incoherent addition of detected powers from all 128 tiles, 
and were written out in the \PSRFITS\ data format \citep{hotan+2004} after summing 
the two linear polarizations.  The resultant spectra have a temporal resolution of 100 $\mu$s and a 
spectral resolution of 10 kHz. 

Fig.~\ref{fig:presto} shows the pulsar detection from our first observation. With a signal-to-noise ratio, 
S/N$\sim$205, the implied mean pulsar flux is $\sim$2 Jy, assuming the nominal MWA sensitivity.
In the second observation the pulsar was significantly weaker (S/N$\sim$80), even though the array had 
$\sim$85\% of its nominal sensitivity;  the change in flux (down to $\sim$1 Jy) is thus primarily due to 
scintillation.

\begin{figure}[t]
\epsscale{0.10}
\includegraphics[width=7.0cm,angle=270]{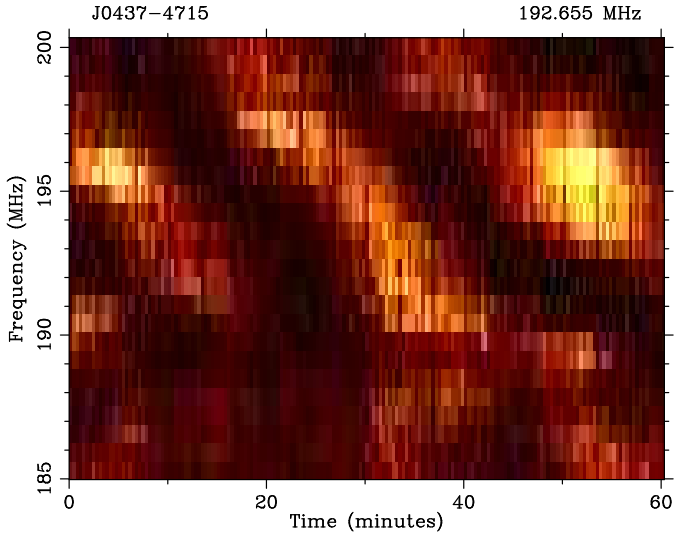}
\caption{Dynamic spectrum of the pulse intensity from the MWA observations of \psr\ over a time 
duration of 1 hr and a bandwidth of 15.36 MHz centred at 192.655 MHz.  The time and frequency 
resolutions in this plot are 20 seconds (i.e. one sub-integration) and 0.64 MHz, respectively.}
\label{fig:dspsr}
\end{figure}

The use of \PSRFITS\ data format provides compatibility with standard pulsar 
software packages, \PRESTO\ (Ransom 2001) and \DSPSR\ \citep{vb2011}.
Initial processing and checks were performed using \PRESTO, following which 
the data were processed using \DSPSR\ to generate synchronously-folded pulse profiles over 
20-second sub-integrations.
These were subsequently averaged in frequency for the improved signal-to-noise needed to 
generate a dynamic spectrum, shown in Fig.~\ref{fig:dspsr}.  
A resolution of 0.64 MHz was chosen to minimise instrumental artefacts that may arise from periodic 
flagging of the edge channels (20 each) on either end of a given coarse (1.28 MHz) channel. 

\section{Results and Discussion} \label{s:res}

\begin{figure}[t]
\epsscale{0.10}
\vskip 0.5cm
\includegraphics[width=6.8cm,angle=270]{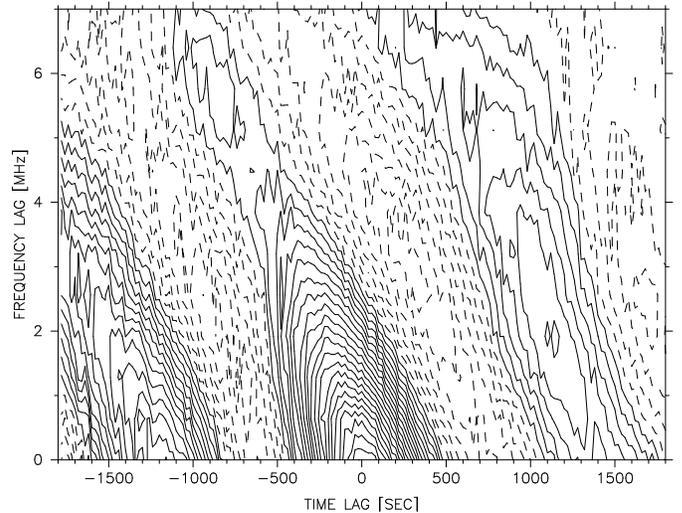}
\caption{Auto-correlation function of the dynamic spectrum of \psr\ shown in Fig.~\ref{fig:dspsr}, 
plotted for lag values out to half the total observing time and half the observing bandwidth.}
\label{fig:acf}
\end{figure}

Our data show rapid, deep modulations of pulse intensity in time and frequency arising from diffractive 
scintillation, as well as prominent drifts of scintillation maxima arising from refractive 
effects caused by discrete, wedge-like density structures. 
Although scintillation observations of \psr\ have been made in the past \citep{nj1995,johnston+1998,gg2000,gwinn2006}, 
this is the first time such drifts have been observed for this 
pulsar. Characteristic scales in time and frequency, and the drift rate in the time-frequency 
plane, can be derived by computing the auto-correlation function (ACF) of the dynamic spectra (Fig.~\ref{fig:acf}). 
The secondary maxima in this plot are due to a few dominant scintles being present in our data (Fig.~\ref{fig:dspsr}). 

Following the published literature \citep{gupta1994,bhat1999,wang+2005}, we fit the ACF with a two-dimensional 
elliptical Gaussian of the form 
$\rho_g\,(\nu,~\tau)=C_0\,{\rm exp}\,[-(C_1\,\nu^2\,+\,C_2\,\nu\,\tau\,+\,C_3\,\tau^2)]$, to yield characteristic 
scales: the decorrelation bandwidth, 
$\nu_d=({\rm ln\,2}/C_1)^{0.5}$, measured as the half-width at half-maximum of the correlation peak; 
scintillation timescale, $\tauiss=(1/C_3)^{0.5}$, measured as the $1/e$ width of the correlation 
function; and the drift rate, 
$dt/d\nu=-(C_2/2\,C_3)$. The presence of significant refractive bending results in an underestimation of 
decorrelation bandwidth as the drifting patterns are no longer aligned in time, however this can be alleviated to a 
certain extent \citep[cf.][]{bhat1999} by the use of a ``drift-corrected" decorrelation bandwidth, \nudc, defined as 
\be 
\nudc\,=\,({\rm\,ln\,2})^{0.5}\,\left(C_1-{C_2 ^2\over4\,C_3}\right)^{-0.5}. 
\ee
The measurement errors on these quantities are largely due to the limited number of scintles in our data, 
given by $\sigma _f ^2\approx(1/f)\,(\tauiss/\Tobs)\,(\nu_d/B)$, where: $f$ is the filling factor 
(nominally assumed 0.2-0.5); $\Tobs$ and $B$ are the net observing time and bandwidth, respectively; and 
$\sigma _f$ is the fractional error. 
For the data shown in Fig.~\ref{fig:dspsr}, we estimate $\nudc\,\sim$\,1.7\,MHz,~$\tauiss\,\sim$\,260\,s,~$\dtnu\,\sim$95\,\driftunit, 
and $\sigma_f\,\sim0.2$, with a net measurement error ($\sigma$) $\sim$25\%, accounting for $\sim15\%$ error from model fits.   

\subsection{Scintillation Measurements}

\subsubsection{Decorrelation bandwidth and scintillation scales}

A summary of all published scintillation measurements for \psr\ is given in Table 1. For meaningful comparisons,  we have also tabulated the scaled values of decorrelation bandwidth and scintillation timescale at the MWA's frequency, \nudmwa and \tauissmwa, respectively. As all measurements including ours are obtained from a single or a few epochs of observation, they are subject to large uncertainties (by factors as much  $\sim$2-3) from refractive scintillation 
\citep[e.g.][]{wang+2005,bhat1999,gupta1994}. 
Our measured \nud\ implies a wavenumber spectral coefficient, $\Cnsq\sim9\times10^{-5}$ \cnsqunits, 
corresponding to a scattering measure, SM$\equiv\int\Cnldl\sim4.5\times10^{-6}$ \smunits. These are the second lowest of all measurements published so far. The lowest values of \Cnsq and SM are reported toward PSR B0950+08 \citep{pc1992} that has a slightly higher DM of 2.96 \dmu and located at  
a larger distance ($280\pm25$ pc) compared to \psr.

Table 1 indicates that our measured \nud\ is highly discrepant with the majority of the published values -- by over an order of magnitude, but 
agrees with the larger scale of scintillation from \citet{gwinn2006}. Further, most published values are, if at all, closer to the smaller scale in their observation.\footnote{Our data do not show any evidence of a smaller scale, though this may be due to insufficient signal-to-noise at high (10 kHz) resolution.}
\citet{gwinn2006} interpret their observations in terms of  ``two scales of scintillation,"  corresponding to two scales of structure.
However, appearance of two distinct scales (in time and frequency), 
separated by $\sim$1-2 orders of magnitude, can occur when two or more scattered sub-images superpose at the observer, giving rise to periodic patterns -- ``interstellar fringes" -- in pulsar dynamic spectra \citep[e.g.][]{rickett+1997,gupta+1999}. 
Furthermore, these are expected to be transitory, with variable widths on time scales ranging from days to months \citep[e.g.][]{rickett+1997}, and require more refraction than that predicted by a pure Kolmogorov medium, particularly for observations at frequencies 
well below the transition frequency ($\nuc$), where decorrelation bandwidth \nud\,$\sim$ the observing frequency, $\nuobs$. Our measured \nud\ implies  \nuc$\sim$1 GHz, and incidentally, all 
the reported scintillation measurements of \psr\ are from observations at frequencies \la600 MHz.

\subsubsection{Scintillation velocity and location of the scattering screen}

Measurements of decorrelation bandwidth (\nud)  and scintillation timescale (\tauiss) can be used to derive the scintillation velocity, \Viss, 
i.e. the net transverse motion between the pulsar, the observer and the medium. 
The measured values  of \nud and \tauiss (and hence \Viss) critically depend on both the line-of-sight and transverse distributions of 
scattering material, as well as  on the wavenumber spectrum of plasma irregularities. 
The transverse distribution is particularly important because the physical extent of screen (or medium) influences the apparent decorrelation
bandwidth; for instance, a finite transverse extent of screen will give rise to a larger value for \nud \citep{cl2001}. 
Moreover, statistical inhomogeneities will lead to time-dependent modulations of \nud and \tauiss, as often observed \citep[e.g.][]{gupta1994,bhat1999};   such effects can be more prominent at low frequencies because of inherently wider scattering cones.

Detailed treatments relating the quantities \nud\ and \tauiss\ to \Viss\ are given by \citet{cr1998} and \citet{dr1998}. Their formalisms can be used to derive improved pulsar distance estimates, if the pulsar proper motion ($\mu$) is also known. In the case of \psr, both the distance and proper motion are very well constrained \citep{verbiest+2008,deller+2008}, and hence the knowledge of \Viss and proper motion, together with the measurements of \nud and \tauiss, can be used to place constraints on location of the scattering screen. For the simplest case, where the scattering medium is approximated as a thin screen located between the pulsar and the observer, \Viss is given by \citep[cf.][]{gupta1994}
\be
\Viss~=~\Aiss\,{\sqrt{\nud\,\Dpsr\,x}\over{\nuobs\,\tauiss}}, 
\label{eq:viss} 
\ee
where \Dpsr\ is the pulsar distance, and $x=\Dos/\Dps$, i.e. the ratio of the distances from the screen to the observer ($\Dos$) and from the screen to the pulsar ($\Dps$). The constant, \Aiss\, relates the decorrelation time (\tauiss) to the velocity, for which \citet{cr1998} derive $2.53\times10^4$ for a Kolmogorov turbulence spectrum and homogeneously-distributed medium, whereas for a single asymmetrically-located thin screen, \citet{gupta1994} derive $\Aiss\,=\,3.85\,\times\,10^4$. The scintillation velocity is essentially a vector addition of the pulsar's transverse velocity, $\Vmu=\mu\,\Dpsr$, to terms such as, (i) the binary transverse motion of the pulsar, \Vbin, (ii) the component of the Earth's orbital velocity, \Vearth, and (iii) an unknown bulk motion of the screen, \Vscreen. Including the pulsar binary motion, eq. (4) of \citet{gupta1995} can be re-written as
\be
\Viss\,=\,|x\,\Vmu+\Vbin+\Vearth-(1+x)\,\Vscreen|.  
\ee
For \psr,  the error on \Viss\ is largely due to measurement errors on \nud and \tauiss.  Even though refractive modulations may alter these quantities, on the basis of a predicted (and observed) positive correlation of their variabilities, we expect the \Viss measurements to be relatively stable.  Since \Vbinm $\sim$13\,\velu , even after accounting for a maximal contribution from the Earth's motion (30\,\velu) and a nominal $\sim$15\,\velu for \Vscreen \citep{bondi+1994}, we estimate \Viss$\sim$325\,\velu, which is on the higher end of the range of values derived from past observations. 

As seen from Table 1, almost all measured values of \Viss\ are consistently larger than the pulsar's transverse velocity, \Vmu$\sim$100\,\velu.  
Based on the formalism and the underlying lever-arm argument in deriving eq.~\ref{eq:viss}, this would strongly suggest a screen location that 
is closer to the pulsar. Specifically, our measurement yields $x=\Viss/\Vmu\sim3$, implying a scattering screen located at $\sim$120\,pc from 
the observer. 
A similar inference was made by \citet{gg2000} based on their observations at 327 MHz (Table 1), suggesting $\Dos\sim0.6\,\Dpsr$. 
Conservatively, $x\sim$1-3 from the published data so far, which would translate to a screen location, \Dos$\sim$(0.5-0.8)\Dpsr, 
i.e. $\sim$80-120\,pc. The formalism of \citet{dr1998}, which considers a more realistic case of a discrete scatterer and a distributed component, predicts \Dos$\sim$(0.65-0.95)\Dpsr, with an implied \thetaHtau$\sim$0.2-0.6, the 
ratio of the source size to the scattering angle. Interestingly, this inferred screen location compares well with the estimated size of the Local Bubble 
in this direction, which is $\sim$100-120\,pc  \citep{bhat+1998}. The bubble extends out to $\sim$10-50\,pc in 
the Galactic plane and $\sim$100-200\,pc perpendicular to the plane \citep[e.g.][]{snowden+1990,bhat+1998,ne2001}. 

This is however in stark disagreement 
with \citet{smirnova2006}, who suggest a location $\sim$10 pc from the Sun. 
Their conclusion relies on a better agreement of their estimated scattering angle, \thetadiff, from the 
observed scintillation timescale of \citet{gwinn2006} (and some reasonable assumptions)  
with the predictions 
based on the statistical dependence of \thetadiff-DM (from \citet{pynzar1997}). 
The empirical relation between scattering and DM is however poorly constrained at low DMs, \la\,10\,\dmu 
\citep[cf.][]{bhat+2004}. Furthermore, their relative refractive angle ($\thetaref\sim2\thetadiff$) 
is inconsistent with that inferred from our observations. For a mid-way location of screen, we estimate $\thetadiff\,\sim0.7$\,mas 
and $\thetaref\,\sim$0.3\,mas, and hence $\thetaref\,\sim\,\thetadiff/2$.
Furthermore, a $\sim$10\,pc screen location would mean $x\sim\Dos/\Dps\sim0.1$, implying very low values of \Viss, which is not supported by any of the observations so far.

\subsection{Pulse profile evolution with radio frequency}

\begin{figure}[b]
\epsscale{0.25}
\vskip 0.5cm
{\hskip -1.8cm
\includegraphics[width=10.5cm,angle=0]{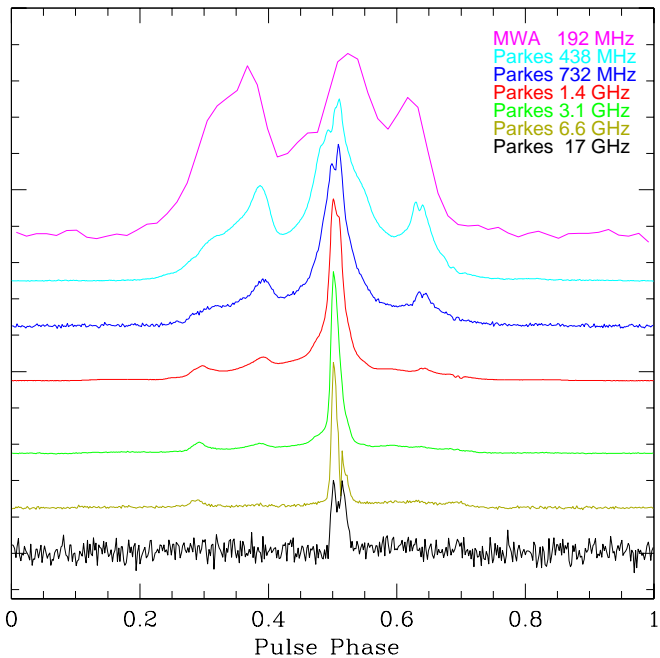}
}
\vskip -0.88cm
\caption{Integrated profiles of \psr\ at frequencies from 200 MHz to 17 GHz.
Archival data from Parkes were processed to generate pulse profiles with 
512 phase bins across the pulse period (i.e. a time resolution $\approx11.23\,\mu$s), whereas the 
MWA profile has 64 phase bins, i.e. approximately 90\,$\mu$s time resolution.}
\label{fig:profs}
\end{figure}

Fig.~\ref{fig:profs} presents the integrated pulse profiles of \psr\ at frequencies from 0.2 to 17 GHz, 
spanning almost two orders of magnitude in frequency.  
Data at 438 MHz are from \citet{navarro+1997}, while those at higher frequencies are from the Parkes 
observatory pulsar data archive \citep{hobbs+2011}. 
At frequencies from 0.4 to 3 GHz where it has been intensely studied, the central bright (core) 
component is flanked by multiple outer (conal-like) components that are also asymmetric, with a 
characteristic ``notch'' clearly visible (in both total intensity and polarisation) in the data 
taken at 0.43 and 1.3 GHz \citep{navarro+1997,dyks+2007}.  
Modelling work  has identified at least seven clear components in its profile, i.e. three distinct 
cones surrounding a central core \citep{gil+1997,qiao+2002}. 
The emission physics and beam modelling is difficult for pulsars in general, and with the 
observed complexity in polarisation and the pulse profile,  it becomes particularly challenging for \psr. 

The pulse profiles were nominally aligned using a simple template matching technique (Taylor 1992) as implemented 
within \PSRCHIVE. 
The peak of the core component is fairly well aligned in this process, except at 17 GHz, where the profile is 
significantly different  and aligned near the centre of the bridge emission.

The pulse shape is evidently complex, with the emission typically spanning $\pm100^{\circ}$ 
in longitude around the core peak (Fig.~\ref{fig:profs}). The relative longitudinal phase shifts of the conal 
components are $\sim$0.05~($20^{\circ}$) between the leftward peaks (at 1.4 GHz and 0.2 GHz) and 
$\sim$0.025 ($10^{\circ}$) between the rightward peaks. These outer components evolve to become almost as strong as 
the core component at 0.2 GHz, indicating their relative spectra are significantly different from those of the central components. 
Further, a mean flux of $\sim$1-2 Jy indicated by our observations may imply a possible spectral break in the $\sim$200-400 MHz 
range \citep[cf.][]{mcconnell+1996}.

The complex pulse shape of \psr\ and its strong evolution with frequency make modelling in 
terms of emission beams challenging. Both the coherent 
curvature radiation \citep{gil+1997} and  inverse compton scattering \citep{qiao+2002} models reproduce the 
observed pulse shapes (at 436 and 1382 MHz) with reasonable success, with an emission geometry comprising core and multiple 
outer cones. However, the asymmetry seen with the outer conal components is not reproduced by either of the 
models. Furthermore, neither model takes into account the relative longitudinal phase shifts of the conal 
components, which \citet{gt2006} interpret in terms of retardation-aberration effects and estimate the emission altitudes to 
be $\sim$10-30\% of the light cylinder radius.

\begin{deluxetable}{lccccccc}[t]
\tablewidth{0pc}
\tablecaption{Scintillation measurements of \psr \label{tab:scnt}}
\tablehead{
\colhead{MJD} & \colhead{\nuobs} &   \colhead{\nud} &  \colhead{\nudmwa$^b$} &  \colhead{\tauiss} &  \colhead{\tauissmwa$^b$} &
\colhead{\Viss} & \colhead{Ref.}\\
 & (MHz) &  (MHz) & (MHz) & (min) & (min) & (${\rm km\,s^{-1}}$) & 
}
\startdata
48825 & 436  & 3.2-4.4 & 0.13-0.18  & 4.6-10.9 &1.7-4.1& 114-288$^a$     & 1 \\
49459 & 436 & 3.6        & 0.15          & 7.8        &2.9     & 170        & 2 \\
          & 660 & 17.4     & 0.14          & 10          &2.3     &              & 2 \\
50523 & 327 & 0.2-3.0 & 0.03-0.38  & 1.8-5.0    &1.0-2.6& 171-321 & 3 \\
50392 & 328 & 16        & 1.98           & 17         &9        & 110-200 & 4 \\
          & 328 & 0.5       & 0.06           & 1.5        &0.8     &             & 4 \\
56559 & 192 & 1.7       & 1                & 4.5        &1       & $325\pm90$      & 5 \\
\enddata
\tablecomments{References: (1) \citet{nj1995}; (2) \citet{johnston+1998}; (3) \citet{gg2000}; (4) \citet{gwinn2006}; (5) This work.\\ $^a$ Re-scaled for $\Aiss=3.85\times10^4$ \citep[cf.][]{gupta1994}, the adopted value for all other Refs.\\ $^b$Scaled to the MWA frequency \nudmwa=192 MHz using a scaling index, $\alpha$=3.9  for 
$\nud\propto(\nuobs)^\alpha$ \citep[cf.][]{bhat+2004} and assuming $\tauiss\propto(\nuobs)^{1.2}$}.
\end{deluxetable}

The MWA detection also highlights the importance of accounting 
for spectral evolution of the pulsar emission in accurate determinations of DMs. 
At low frequencies, frequency-dependent DM changes may also arise from multi-path 
propagation effects, as the radiation at different frequencies samples slightly different 
total electron contents (\thetadiff$\propto\nuobs^{-2.2}$). 
Any unmodelled profile evolution may thus manifest as DM variations, as demonstrated by 
\citet{hassall+2012} and \citet{ahuja+2007}, albeit for long-period pulsars. 
Recent work of \citet{pennucci+2014} involving a two-dimensional template portrait is 
very promising in this context.

DM variations $\sim10^{-3}$\,\dmu have been seen toward \psr\ over $\sim$5-10\,yr, and even larger variations ($\sim10^{-2}$\,\dmu) toward other pulsars \citep{keith+2013}. For typical DM accuracies achievable with the PPTA data ($\sim10^{-4}$\,\dmu), the differential delay across the full MWA frequency range is $\approx60\,\mu$s. The large frequency lever arm possible with the MWA, e.g., by spreading out $24\times1.28\,{\rm MHz}$ channels of VCS non-contiguously over the 220 MHz band, so as to simultaneously sample multiple spot frequencies within the 80-300 MHz range, can be exploited to enable such wide-band observations. 

\section{Future prospects}\label{s:conc}

With its sensitivity, field-of-view and frequency coverage, the MWA makes a major facility for low-frequency pulsar astronomy. 
Our current capabilities will soon be boosted with the implementation of the full-bandwidth recording and phased-array modes, 
resulting in an over an order-of-magnitude improvement in sensitivity. 
The combination of field-of-view and VCS functionality can be exploited for realising observations of multiple timing-array 
pulsars from a given pointing, and the flexibility to spread out the 30.72 MHz bandwidth anywhere within the 80-300 MHz range 
can be leveraged for observations at multiple frequencies simultaneously.\\

\noindent{\it Acknowledgements:}
We thank the referee for several insightful comments that helped improve the paper. 
This scientific work makes use of the Murchison Radio-astronomy Observatory, operated by CSIRO. We acknowledge the Wajarri Yamatji people as the traditional owners of the Observatory site.  NDRB is supported by a Curtin Research Fellowship. 
We thank Matthew Bailes and Bryan Gaensler for useful comments, and Dick Manchester and Lawrence Toomey for help with access to the 430 MHz archival data. 
Support for the MWA comes from the U.S. National Science Foundation (grants AST-0457585, PHY-0835713, CAREER-0847753, and AST-0908884), the Australian Research Council (LIEF grants LE0775621 and LE0882938), the U.S. Air Force Office of Scientific Research (grant FA9550-0510247), and the Centre for All-sky Astrophysics (an ARC Centre of Excellence funded by grant CE110001020). Support is also provided by the Smithsonian Astrophysical Observatory, the MIT School of Science, the Raman Research Institute, the Australian National University, and the Victoria University of Wellington (via grant MED-E1799 from the New Zealand Ministry of Economic Development and an IBM Shared University Research Grant). The Australian Federal government provides additional support via CSIRO, National Collaborative Research Infrastructure Strategy, Education Investment Fund, and the Australia India Strategic Research Fund, and Astronomy Australia Limited, under contract to Curtin University. We acknowledge the iVEC Petabyte Data Store, the Initiative in Innovative Computing and the CUDA Center for Excellence sponsored by NVIDIA at Harvard University, and the International Centre for Radio Astronomy Research , a joint venture of Curtin University and The University of Western Australia, funded by the Western Australian State government.



\bibliographystyle{apj}

\end{document}